\documentstyle[12pt]{article}
\newcommand{\be}{\begin{equation}}
\newcommand{\ee}{\end{equation}}
\newcommand{\ba}{\begin{eqnarray}}
\newcommand{\ea}{\end{eqnarray}}

\begin{document}
\begin{center}
{\bf  MATRIX HAMILTONIANS: SUSY APPROACH TO
 HIDDEN SYMMETRIES}\\

\vspace{1cm}
A. A. ANDRIANOV\footnote{{\it Department of Theoretical
Physics, University of Sankt-Petersburg,198904 Sankt-Petersburg, Russia.
E-mail:} andrianov1@phim.niif.spb.su;\hspace{3ex}
ioffe@phim.niif.spb.su},
F. CANNATA\footnote{{\it Dipartimento di Fisica and INFN, Via Irnerio 46,
40126 Bologna, Italy. E-mail:} cannata@bo.infn.it},
 M. V. IOFFE$^a$,  D.N.NISHNIANIDZE$^{a,}$\footnote{\it On leave of absence
from Kutaisi Polytechnic University, Georgia}
\end{center}
\newpage
\vspace{1cm}
\hspace*{0.5in} \begin{minipage}{5.0in}
{\small
A new supersymmetric approach to the analysis of dynamical
 symmetries for matrix
quantum systems is presented. Contrary to
standard one dimensional quantum mechanics where there is no role
for an additional symmetry due to nondegeneracy, matrix
hamiltonians allow for non-trivial residual
symmetries.
 This approach is based on a generalization of
the intertwining relations familiar in SUSY Quantum Mechanics.
 The corresponding matrix supercharges,
of first or of second order in derivatives, lead to an algebra
which incorporates an additional
block diagonal differential matrix operator (referred to as a "hidden"
 symmetry operator) found to commute with
the superhamiltonian.
We discuss some physical interpretations
of such dynamical systems in terms of spin $1/2$ particle in a magnetic field
or in terms of coupled channel problem. Particular attention is
paid to the case of transparent
matrix potentials.}

\end{minipage}
\newpage
\section*{\large\bf 1.\quad Introduction}
\hspace*{3ex} Supersymmetric Quantum Mechanics (SUSY QM) \cite{witten}
is an interesting framework
to analyze non-relativistic quantum problems. In particular it allows to
 investigate the spectral properties of
certain quantum models as well as to generate new systems with given
spectral characteristics.

 In general it is well known that SUSY algebra
provides the relation between (super)partner hamiltonians
 (which are often referred as "bosonic" and
"fermionic")  associated to
a two-fold degeneracy of levels.

 Much less attention has been paid in the
literature to the possibility of using SUSY as a tool to study
individual ("internal") symmetries of each superpartner hamiltonian.
In principle this can be useful if it allows to  integrate partially
dynamical systems by discovering an additional dynamical
symmetry which we shall refer to as a "hidden" symmetry.
We will show that indeed hidden symmetries may be revealed by SUSY inspired
approaches \footnote{Notice that the existence of a
symmetry operator is not
necessarily associated with degeneracy of
eigenstates of the Hamiltonian. One can convince
oneself of this fact for matrix Quantum Mechanics
taking $2\times2$ diagonal hamiltonian with
components having different spectra and no degeneracy.
In this case a symmetry operator is obviously $\sigma_3$ playing
the role of a grading operator.}.

The standard SUSY QM relations read
\be
\{Q^+, Q^-\} = H = \left(\begin{array}{cc}
h^{(1)} & 0 \\ 0 & h^{(2)} \end{array}\right) =
\left(\begin{array}{cc}
q^+q^- & 0 \\ 0 & q^-q^+ \end{array}\right);  \label{hamilt}
\ee
$$h^{(i)} \equiv - \partial^2 + V^{(i)} (x); \quad q^{\pm} =
\mp \partial + W(x); \quad \partial \equiv \partial/\partial x; $$
\be
Q^{-} = \left( \begin{array}{cc}
0 & 0 \\ q^- & 0 \end{array} \right)
\hspace{1cm} \hspace{1cm}
Q^{+} = \left( \begin{array}{cc}
0 & q^{+} \\ 0 & 0 \end{array} \right), \label{largeQ}
\ee
\be
\{ Q^{\pm},Q^{\pm} \} = 0. \label{anticommut}
\ee
\be
[ Q^{\pm},H ] = 0; \qquad
h^{(1)}q^+ = q^+h^{(2)} \label{commut}
\ee

In general there are different realizations of this algebra,
e.g. multidimensional
\cite{abi} and matrix ones \cite{ci}. It is also possible to generalize
the algebra itself
by preserving Eqs. (\ref{anticommut}), (\ref{commut}) and
allowing, by a non standard form
of the intertwining operators $q^{\pm},$ to modify (\ref{hamilt}) to
become
\be
\{Q^+, Q^-\} = K = \left(\begin{array}{cc}
k^{(1)} & 0 \\ 0 & k^{(2)} \end{array}\right) =
\left(\begin{array}{cc}
q^+q^- & 0 \\ 0 & q^-q^+ \end{array}\right),\label{quasiham}
\ee
where the diagonal operator $K$ is no more the
superhamiltonian but has in general
the nature of a symmetry operator
\be
[ k^{(i)},h^{(i)} ] = 0. \nonumber
\ee

This generalization was already discussed for 1-dim \cite{ais}
and 2-dim \cite{ain}
Quantum Mechanics. For 1-dim systems and intertwining operators
of second order in
derivatives the only relevant case was the one for which $K$ is
a function of $H,$
$$
K = H^2 - 2 \alpha H + \beta $$
where $\alpha$ and $\beta$ are constants.
For the 2-dim case there exists the possibility of having a central
charge $R,$ which
commutes with all elements of algebra, such that
\be
K = f(H) + R. \label{subtr}
\ee
 A new, supersymmetrical, method was elaborated
\cite{ais}, \cite{ain} to investigate
hidden dynamical symmetries of quantum systems.
The existence of such a differential operator $R$ implies a
dynamical symmetry (unknown a priori)
made apparent by the generalized SUSY algebra.

Let us remind the physical impact of the supersymmetric approach to the Dirac
equation \cite{kostel} with some applications to superconductivity
\cite{moreno},
to pseudorelativistic behaviour of electrons in two-band systems
\cite{ferrari} and to attempts to a
diagonalization procedure\cite{ci}) in nuclear and atomic physics
coupled channel problems and finally
in the treatment of particles with spin
in an external magnetic fields \cite{ai}. Therefore it is
important to discuss the role of symmetry operators
for 1-dim dynamical systems of Schr\"odinger type with
matrix potentials. Formally the most
straightforward method is to study the commutator
of the Hamiltonian matrix with a generic
differential operator matrix $R$ and solve the corresponding
system of differential
equations obtained by imposing the commutator to vanish $[h , R] = 0.$
Incidentally one can easily check that for the scalar case
this system of equations
(after possible subtraction of the Hamiltonian) does not allow
for non-trivial solution.
On the contrary in the matrix case non-trivial solutions exist
even for the case of symmetry
operators of first order in derivatives. For higher
order derivatives the equations become
rather cumbersome and it is too difficult to provide the
general discussion of the solutions.

For this higher order case a  method of solution suggested by
supersymmetry seems to be useful.
It starts from the same idea of factorization originally proposed by
Schr\"odinger \cite{schroed}
with the related intertwining operators (see (\ref{commut}))
but now it is not applied to the Hamiltonian but rather
to the symmetry operator:
 this method reduces the order of differential equations one must
solve without
however increasing their number.
In this paper we shall find non-trivial genuine second order
operators $R$ even for 1-dim first order intertwining matrix operators.

The paper is organized as follows. In Section 2 we shall
investigate the SUSY approach with
intertwining operators of first order in derivatives. A variety of
matrix systems allowing for
the genuine second order symmetry operators will be obtained.
In the Section 3 this method will be applied to a more complex case
of higher order intertwining operators, examining in particular
the conditions for the
factorizability of these intertwining operators (reducibility).
The type of physical systems which we can describe in our formalism include
a particle with magnetic moment in a magnetic field and more generally
a class of systems with one continuous and one discrete degrees of freedom.
A novel construction is given for transparent matrix potentials
which are not
duplications of standard scalar transparent potentials and are not generated
by iterations of first order Darboux transformations.

\section*{\large\bf 2.\quad First order Matrix SUSY Quantum Mechanics }
\hspace*{3ex}
We start from the general first order (in derivatives)
representation of the components
of supercharges in the case of 1-dim Quantum Mechanics
\be
q^+ \equiv A \partial + \widetilde B ; \quad
q^- \equiv - A^{\dagger} \partial + \widetilde B^{\dagger},  \label{qpm}
\ee
where $A$ and $\widetilde B$ are matrices. Imposing that
these operators intertwine the Hamiltonians
$h^{(1)}$ and $h^{(2)}$ (see Eq.(\ref{commut})) reads
\be
(- \partial^2 + V^{(1)}(x)) (A \partial + \widetilde B) = (A \partial
+ \widetilde B) (- \partial^2 + V^{(2)}(x)),
\label{intertw}
\ee
where $V^{(1)}, V^{(2)}$ are hermitean potential matrices.
Eq.(\ref{intertw}) amounts to solve the
following three equations:
\be
A^{\prime} = 0; \label{intertw1}
\ee
\be
V^{(1)} A - A V^{(2)} = 2 \widetilde B^{\prime}; \label{intertw2}
\ee
\be
- \widetilde B^{\prime \prime} + V^{(1)} \widetilde B -
\widetilde B V^{(2)} - A V^{(2) \prime} = 0. \label{intertw3}
\ee
The first equation implies that  from now on we will assume $A$
to be a constant matrix.
As usual in the frameworks of SUSY QM the intertwining relations
Eq.(\ref{intertw}) lead to the
connection between the column eigenfunctions of Hamiltonians
$h^{(1)}$ and $h^{(2)}:$
\be
\Psi^{(1)}(x) = (A \partial + \widetilde B ) \Psi^{(2)}(x); \quad
\Psi^{(2)}(x) = (A \partial + \widetilde B )^{\dagger} \Psi^{(1)}(x)
\label{wavefunc}
\ee
The Eq.(\ref{wavefunc}) allows sometimes for zero-modes, and
the spectra of the partner hamiltonian coincide
up to these zero-modes.

What is not standard is the fact that Eq.(\ref{hamilt}) does not hold
because the products of the supercharge
components are no more equal to the Hamiltonians:
\be
q^+ q^- = k^{(1)} =
-A A^{\dagger} \partial^2 + (A \widetilde B^{\dagger} -
\widetilde B A^{\dagger}) \partial +
\widetilde B \widetilde B^{\dagger} + A
\widetilde B^{\dagger \prime}; \label{k^1}
\ee
\be
q^- q^+ = k^{(2)} =
- A^{\dagger} A \partial^2 + ( \widetilde B^{\dagger} A -
A^{\dagger} \widetilde B ) \partial + \widetilde B^{\dagger}
\widetilde B - A^{\dagger} \widetilde B^{\prime}. \label{k^2}
\ee

It is natural to consider separately $\mathop{det} A \not = 0$
and $\mathop{det} A = 0$
because the way to solve the system of equations
(\ref{intertw1}), (\ref{intertw2}), (\ref{intertw3})
differs for the two cases.
In the case $\mathop{det} A \not = 0$ it is possible to subtract the
Hamiltonians $h^{(1)}$ and $h^{(2)}$ from $k^{(1)}$ and $k^{(2)}$,
being left with a symmetry operators
of first order in derivatives; otherwise we deal with second order operators.
Furthermore by a suitable similarity transformation induced by the matrix $A$
itself it is possible to consider $A = -1$ as a representative
of the case $\mathop{det} A \not = 0$
whereas we can take the matrix
$A = \left( \begin{array}{cc} a & b \\ 0 & 0 \end{array} \right)$
as representative of $\mathop{det} A = 0.$

\subsection*{\normalsize 2.1.\quad The case $A = -1$, symmetry
operators of first order}
\hspace*{1ex}
In this case Eqs.(\ref{intertw2}) and (\ref{intertw3}) become
\be
V^{(2)} - V^{(1)} = 2 \widetilde B^{\prime}; \label{intertw4}
\ee
\be
- \widetilde B^{\prime \prime} + V^{(1)} \widetilde B -
\widetilde B V^{(2)} +
V^{(2) \prime} = 0. \label{intertw5}
\ee
One can parametrize the potential matrices taking into account
(\ref{intertw4}):
\be
V^{(1)} = {\widetilde B}^2 -  {\widetilde B}^{\prime} + W; \quad
V^{(2)} = {\widetilde B}^2 + {\widetilde B}^{\prime} + W. \label{parametr}
\ee
Then Eq.(\ref{intertw5}) becomes simply
\be
W^{\prime}(x) = [ \widetilde B(x), W(x) ]. \label{simple}
\ee
Whoever is familiar with ordered products, e.g. in Gauge
Field Theory, will not have
difficulty to recognize that Eq.(\ref{simple})
has not a simple local solution unless
$$[\widetilde B(x), \widetilde B(y)] = 0.$$
Since effectively this last condition reduces the matrix problem to a
scalar problem we do not
find this case of any interest for us.

>From (\ref{intertw4}) one deduces that the non-hermitean
part of $\widetilde B$ does not depend on $x$ because
of the hermiticity of the potentials. Therefore we can
parametrize  $\widetilde B(x) = B(x) + iC,$ where $B$
and $C$ are both hermitean matrices and  $C$ is a constant matrix.
Correspondingly one has to solve
the  system of matrix equations
\be
W^{\prime} = [B,W] + i[C,W]; \label{intertw6}
\ee
\be
W - W^{\dagger} = -2i\{ B,C \} \label{hermit}
\ee
which are the consequence of (\ref{intertw5}) and of the hermiticity
of both potentials.
Restricting to $2 \times 2$ matrix potentials
one can study the system of equations
(\ref{intertw6}) and (\ref{hermit}) by
expanding all matrices in Pauli
matrices and unity:
\be
B = b_0 + b_i \sigma_i ; \quad C = c_0 + c_i \sigma_i; \quad W =
w_0 + w_i \sigma_i , \label{expansion}
\ee
with components $b_0, b_i$ and $c_0, c_i$ real. The related
symmetry operators
read:
\be
R^{(1)} = q^+ q^- - h^{(1)} = 2iC \partial + 2 C^2 - 2i BC - W;
\label{hidden1}
\ee
\be
R^{(2)} = q^- q^+ - h^{(2)} = 2iC \partial + 2 C^2 - 2i CB + W.
\label{hidden2}
\ee

While it is not restrictive
to set $c_1=c_2=0 ,$ it is not interesting
to choose also $c_3$ and $c_0$ both to vanish. Indeed in this case
the symmetry operators above are no more differential
operators and become proportional to a constant matrix $W.$

A general solution of the nonlinear system of matrix equations
(\ref{intertw6}), (\ref{hermit}) amounts first to find a solution
of the subsystem:
\ba
Im w_0 &=& -2 (b_0c_0 + b_3c_3); \nonumber \\
Im w_1 &=& -2 b_1c_0; \nonumber \\
Im w_2 &=& -2 b_2c_0; \nonumber \\
Im w_3 &=& -2(b_3c_0 + b_0c_3); \nonumber \\
w_0 &=& const; \nonumber \\
Re w_3 &=& const, \nonumber
\ea
however a complete solution of
(\ref{intertw6}), (\ref{hermit}) cannot be written.
Particular solutions of (\ref{intertw6}) and (\ref{hermit}) in terms of the
components (\ref{expansion}) can be obtained by making specific
ans\"atze:

1) Let us assume
$$c_0 = b_3 = Re w_3 = 0; \quad Im w_3(x) \equiv 0;  \quad c_3 \not = 0, $$
$ b_3 \not = 0$ would correspond to a trivial solution $W(x) = Const .$
Then one obtains $$b_0 = 0, \, Re w_1 = \beta \cos(2c_3x), \,
Re w_2 = -\beta \sin(2c_3x), \, b_2(x) = -b_1(x)\tan(2c_3x), $$
where $\beta$ is a constant parameter and $b_1(x) \equiv \tilde b_{1}(x)
\cos(2c_{3}x)$ with
$\tilde b_{1}(x)$ being an arbitrary nonsingular function.
Correspondingly the potentials read
\ba
V^{(1),(2)}(x) &=& \tilde b_1^2(x) - c_3^2 + Re w_0\nonumber\\
&+& \bigl[ \beta\cos(2c_3x) \mp \tilde b_1^{\prime}(x)\cos(2c_{3}x) \pm
2c_{3}\tilde b_{1}\sin(2c_{3}x)
 \bigr]\cdot \sigma_1 \nonumber\\
&-&
\bigl[ \beta\sin(2c_3x) \mp \tilde b_1^{\prime}(x)\sin(2c_3x)
\mp 2c_{3}\tilde b_{1}(x)\cos(2c_{3}x) \bigr]\cdot
\sigma_2   \label{poten1}
\ea
and the symmetry operators (\ref{hidden1}), (\ref{hidden2}) after
subtraction of constants become:
\be
R^{(1),(2)} = i\sigma_3\partial \mp
\bigl[ \tilde b_1(x)\sin(2c_3x) + \frac{\beta}{2c_3}\cos(2c_3x)
 \bigr]\cdot \sigma_1
\mp
\bigl[ \tilde b_1(x)\cos(2c_{3}x) - \frac{\beta}{2c_3}\sin(2c_3x)
 \bigr]\cdot\sigma_2 .   \label{symm1}
\ee

Among the different interpretations concerning the physics of
the matrix potentials $V^{(1),(2)}$
one consists of a spin $1/2$ neutral
particle in a (inhomogeneous) magnetic field.
It is necessary to assume that the magnetic field depends only on the
coordinate $x \equiv x_3$ and lies in
the $(x_1, x_2)$ plane in order to ensure the vanishing of
its divergence $\partial_i B_i = 0$.
While the motion in the
$(x_1, x_2)$ plane is trivially free the dynamics is
still rather interesting because of the $x_3$ motion \cite{ai}.
Physicswise the inhomogeneity of the magnetic field is determined in this
case by the requirement of vanishing of the scalar
potential in (\ref{poten1}) (neutrality of the particle).

2) Let us assume
$$c_0 = b_3 = Re w_3 = 0; \quad Im w_3(x) \not = 0  \quad c_3 \not = 0, $$
as before $ b_3 \not = 0$ would correspond to a trivial
solution $W(x) = Const .$
Then one obtains $$b_0(x) = -\frac{1}{2c_3} Im w_3(x); \,
b_1(x) = \frac{(Re w_2)^{\prime} + 2c_3Re w_1}{2Im w_3}; \,
b_2(x) = \frac{-(Re w_1)^{\prime} + 2c_3Re w_2}{2Im w_3},
$$
where
$$
Re w_1(x) = \sqrt{(Im w_3)^2 + \beta} \cos f(x) = \sqrt{(2b_0c_3)^2 +
\beta} \cos f(x) ,
$$
$$
Re w_2(x) = \sqrt{(Im w_3)^2 + \beta} \sin f(x) = \sqrt{(2b_0c_3)^2 +
\beta} \sin f(x).
$$
$\beta$ is a constant parameter as well as $c_3$ and $f(x)$ and $Im w_3(x)$
are arbitrary functions.
The potentials  and the symmetry operators read:
\ba
V^{(1),(2)}(x) &=& b_0^2 \mp b_0^{\prime} + b_1^2 + b_2^2 - c_3^2 +
Re w_0 \nonumber\\
&+& \bigl[ 2b_0b_1 + Re w_1 \mp b_1^{\prime} \bigr]\cdot\sigma_1 +
\bigl[ 2b_0b_2 + Re w_2 \mp b_2^{\prime} \bigr]\cdot\sigma_2
\label{poten2}
\ea
\be
R^{(1),(2)} = i\sigma_3\partial \pm
\bigl[ b_2(x) - \frac{Re w_1}{2c_3} \bigr]\cdot \sigma_1
\pm
\bigl[ -b_1(x) - \frac{Re w_2}{2c_3} \bigr]\cdot
\sigma_2 .   \label{symm2}
\ee
In terms of the "magnetic" interpretation given above one can
now notice that the absence of the scalar potential in
(\ref{poten2}) is less restrictive because the magnetic field still
depends on one arbitrary function.
The intrinsic "periodicity" of the
magnetic field  forces a similar periodicity of
the wave function. We warn however not to interpret the periodicity
too naively since it depends in
general on the properties of the arbitrary function $f(x),$
e.g. asymptotically constant magnetic field can be incorporated
in this scheme.

 One can also find solutions for other ans\"atze like e.g.
$$ c_0=0, \, c_3 \not = 0, \, Re w_3 = const \not =0 ,$$
similarly to 1), 2).

\subsection*{\normalsize 2.2.\quad The case $\det A = 0$,
the symmetry operators of second order}
\hspace*{1ex}
The constant
(see (\ref{intertw1})) matrix $A$ has now the form
\be
A = \left( \begin{array}{cc}
a & b \\ 0 & 0 \end{array} \right) \nonumber
\ee
We write the matrices
$V^{(1),(2)}$ and $B$ explicitly and
for simplicity we assume them to be real:
\be
V^{(i)} = \left( \begin{array}{cc}
v^{(i)}_1 &
v^{(i)} \\
v^{(i)} &
v^{(i)}_2 \end{array} \right)
\nonumber
\ee
\be
B(x) = \left( \begin{array}{cc}
b_1 & b_2 \\ b_3 & b_4 \end{array} \right)
\nonumber
\ee

Equations (\ref{intertw2}) and  (\ref{intertw3})
can be solved and the potentials
can be written in a similar way as before in
terms of Pauli matrices, however the reality
condition forces the absence of a $\sigma_2$ term
\footnote{The interpretation of the dynamics as a magnetic
interaction of $s=1/2$ particle is now still possible
but we can also consider it as a coupled
channel problem as discussed in \cite{ci}.}.
The hidden symmetry operators are now of second order in derivatives.

In the case $b \not = 0$ it is possible to find several solutions
dependent on arbitrary functions; in the case  $b = 0, \, a=1$ the
form of the
solutions simplifies. One solution is given by:
\ba
v^{(1)}_1(x) &=& b_1' + b_2^2 + c^2b_2^{-2} + c_1; \qquad
v^{(1)}(x) = -2c b_2' b_2^{-2};\nonumber \\
v^{(1)}_2(x) &=& - \biggl( \frac{b_2'}{b_2} \biggr)' +
\biggl( \frac{b_2'}{b_2} \biggr)^2 + \frac{2b_2'b_1}{b_2} + b_1^2 +
b_2^2 + \frac{c^2}{b_2^2} - b_1' + \tilde{c}_1; \nonumber\\
v^{(2)}_1(x) &=& -b_1' + b_2^2 + \frac{c^2}{b_2^2} + c_1; \qquad
v^{(2)}(x) = - 2b_2'; \label{(a)}\\
v^{(2)}_2(x) &=& \frac{b_2''}{b_2} +
 \frac{2b_2'b_1}{b_2} + b_1^2 +
b_2^2 + \frac{c^2}{b_2^2} - b_1' + \tilde{c}_1\nonumber
\ea
where $b_1(x), b_2(x)$ are arbitrary functions, $b_3(x) =
c\cdot b_2^{-1}(x)$
and $b_4 = 0 .$
The symmetry operators
$R^{(1),(2)}$
can be straightforwardly derived according to (\ref{k^1}),  (\ref{k^2})
and are second order differential operators.

For the second solution $v_1^{(1),(2)}(x)$ and $v^{(1),(2)}(x)$ are the
same as in (\ref{(a)}) and
\ba
v_2^{(1)}(x) = \frac{b_3''}{b_3} -  \frac{2b_3'b_1}{b_3} -
\frac{2b_2'b_4}{b_3} - b_1' + \sum_{k=1}^{4}b_k^2 + \tilde c_1 ;\nonumber\\
v_2^{(2)}(x) = \frac{b_2''}{b_2} + \frac{2b_2'b_1}{b_2} +
\frac{2b_3'b_4}{b_2} + b_1' + \sum_{k=1}^{4}b_k^2 + \tilde c_1 ,
\nonumber
\ea
where $b_2(x), b_3(x)$ are arbitrary functions such that
$b_2\cdot b_3 \not = 0 ,$  $b_4 = const \not = 0 $ and
$$
b_1(x) = (b_2b_3)^{-1} \biggl[ \frac{(b_2b_3)^2}{2b_4} -
(2b_4)^{-1}(b_2^2 + b_3^2) + \alpha \biggr]
$$
with $\alpha$ a constant parameter.

\section*{\large\bf 3.\quad Second order Matrix SUSY Quantum Mechanics }
\hspace*{3ex}
Let us define the second order differential operators
\be
q^+  = (q^-)^{\dagger}
= \partial^2 - 2 F(x) \partial + B(x). \label{gena+}
\ee
\be
q^-  = (q^+)^{\dagger}
= \partial^2 + 2 F^{\dagger}(x) \partial + B^{\dagger}(x) +
2 F^{\prime \dagger}
(x). \label{gena2+}
\ee
where $F(x)$ and $B(x)$ are $2 \times 2$ matrices. This representation for
$q^{\pm}$ can be inserted
into (\ref{largeQ}) ---
(\ref{quasiham}).

The intertwining relations are equivalent to a system of
three non-linear matrix differential
equations:
\be
V^{(1)}-V^{(2)}+4F^{\prime}= 0 \label{eq1}
\ee
\be
F^{\prime\prime}-V^{(1)}F+FV^{(2)}-B^{\prime}-V^{(2) \prime}=0 \label{eq2}
\ee
\be
B^{\prime\prime}+V^{(2) \prime\prime}-V^{(1)}B+BV^{(2)}-2FV^{(2)\prime}=0
\label{eq3}
\ee
Our attitude towards
the solution of this system of equations is that
we consider $ q^{\pm}, h^{(2)},
h^{(1)}$ to be essentially unknown except for Schr\"odinger form
of hamiltonians and assumption of
structure (\ref{gena+}) of the supercharges, so the problem is to
find the solution in terms of the matrices $F(x), B(x), V^{(i)}(x)$.

Due to the
complexity of the problem ( matrix, 2nd derivatives, non-linearity )
it does not
seem realistic to search for a general solution in analytic form,
instead we believe that techniques of Higher Order SUSY QM
as developed in \cite{ais} can
provide a useful tool for solving in an "indirect" way by the ansatz of
factorizability of $q^{\pm}$, i.e. restricting to the {\bf reducible} matrix
Higher Order SUSY QM.

Another possibility which we mention is a particular solution for
which the terms $FV^{(2)}-V^{(1)}F$ appearing in Eq.(\ref{eq2})
reduce to $\{F,F^{\prime}\}$ with the aim of a "direct" integration of
this equation. A sufficient condition which allows this integration is
$$V^{(1)}+V^{(2)}=2P(F)$$
where $P$ is an arbitrary "scalar" function of the matrix $F(x)$ like
for example $P=\Sigma c_n(x)F^n(x)$ where $c_n$ are scalar functions.

\subsection*{\normalsize\bf 3.1.\quad Reducible Higher Order Matrix SUSY QM}
\hspace*{1ex}
A specific ansatz consists in the factorizability of the operators
$q^{\pm}$ of (\ref{gena+}), (\ref{gena2+}) in terms of
ordinary superpotentials  $W(x)$ and $\widetilde W(x):$
\be
q^+ = q_1^+ q_2^+ = (-\partial + W(x))\, (-\partial + \widetilde W(x)),
\label{facta+}
\ee
connected by the ladder equation
\be
q_1^- q_1^+ = q_2^+ q_2^- + \hat \Delta
\hspace{1cm} \mbox{ or } \hspace{1cm}
 W' + W^2 = - \widetilde W'
+ \widetilde W^2 + \hat \Delta , \label{ladder}
\ee
with $\hat \Delta$ a constant hermitean matrix as will be clear
later on.
Let us assume furthermore $\hat \Delta$ to be diagonal\footnote{
If it is not it can be diagonalized by a constant unitary transformation
which also affects other operators.}.
Then $F$ and $B$ of Eqs.(\ref{gena+}), (\ref{gena2+}) are determined
by the superpotentials $W(x), \widetilde W(x):$
\be
2F=W + \widetilde W ,\qquad B=W \widetilde W - \widetilde W^{\prime}
\nonumber
\ee

The factorization Eq.(\ref{facta+}) arises from two successive
standard SUSY QM transformations
\be
\left(\begin{array}{cc}
h^{(1)} & 0 \\ 0 & h \end{array}\right) = \left(\begin{array}{cc}
q_1^+ q_1^- & 0 \\ 0 & q_1^- q_1^+ \end{array}\right)\label{inth1}
\ee
and
\be
\left(\begin{array}{cc}
h & 0 \\ 0 & h^{(2)}+\hat \Delta \end{array}\right) =
\left(\begin{array}{cc}
q_2^+ q_2^-+ \hat \Delta & 0 \\ 0 & q_2^- q_2^+ + \hat \Delta
\end{array}\right)
\label{inth2}
\ee
by deleting the "intermediate" hamiltonian :
\be
H = \left(\begin{array}{cc}
q_1^+ q_1^- & 0 \\ 0 & q_2^- q_2^+ + \hat \Delta\end{array}\right).
\nonumber
\ee

The matrix $\hat \Delta$  has to be  such that $[H,Q^{\pm}] = 0$
and consequently $[q_2^{\pm},\hat \Delta]=0$
which makes clear the reason why $\hat \Delta$ has to be constant
and that $[\widetilde W ,\hat \Delta]=0$.
We therefore can conclude that $[h^{(2)}+\hat \Delta,\hat \Delta]=0$
and this
allows us to identify $R^{(2)}\equiv\hat \Delta$ as a
{\bf symmetry operator} for
$ h^{(2)}.$ If it further commutes with $ h^{(1)}$ this operator
is such that
\be
\{Q^+, Q^-\}  = (H)^2 -\hat\Gamma  \cdot H
\nonumber
\ee
where $\Gamma$ is a block diagonal matrix
\be
\Gamma \equiv \left(\begin{array}{cc}
\hat \Delta & 0 \\ 0 & \hat \Delta \end{array}\right),
\nonumber
\ee
and thus it corresponds to a rather "trivial" $R$ operator.
We shall from now on exclude such a case imposing that the operator
\be
R^{(1)}=-q_1^+ \hat \Delta q_1^- \label{R1}
\ee
be non-trivial \footnote{In the concluding Section
of \cite{abi} Eq. (\ref{R1}) first appeared.}. The last case is
rather interesting because
it incorporates the possibility of having genuine partner symmetry operators
of different orders in the
derivatives.

In order to derive a solution we expand the previous operators
in terms of Pauli matrices:
\be
W(x) = w_0 + w_i\sigma_i; \quad \widetilde W(x) = \tilde w_0 +
\tilde w_3 \sigma_3; \quad
\hat \Delta = \delta_0 + \delta_3 \sigma_3; \quad \delta_3 \not = 0 .
\nonumber
\ee

To illustrate the techniques involved we give now an example:
\subsection*{\normalsize Example 1. \quad $\delta_0 = 0;\quad w_0(x) =
\tilde w_0(x); \quad w_2(x) = 0$}
\hspace*{1ex}
Eq.(\ref{ladder}) takes then the form
\ba
2w_0w_3+w_3^{\prime}-2w_0\tilde w_3+\tilde w_3^
{\prime}-\delta_3=0\nonumber\\
2w_0w_1+w_1^{\prime} =0\label{ex2}\\
w_3^2+w_1^2+2w_0^{\prime}- \tilde w_3^2=0\nonumber
\ea
with the following solution
\be
w_0= \tilde w_0=1/(2x),\qquad w_1=1/x
,\qquad
w_3= \tilde w_3= \delta_3x/2 .
\nonumber
\ee
In order to arrive to an interpretation of these results we have to write
the potentials:
\be
V^{(1)}(x)=W^2-W^{\prime}=7/4x^2+\delta_3^2x^2/4
+ (2/x^2)\cdot\sigma_1 ;
\nonumber
\ee
\be
V^{(2)}(x)=W^2+W^{\prime}+\hat \Delta
=-1/(4x^2)+\delta_3^2x^2/4+2\delta_3\cdot\sigma_3 .
\nonumber
\ee
e.g. eigensolutions
These potentials contain centrifugal singularities and therefore we
pose the eigenvalue problem on the semiaxis
$ x > 0 . $ The physical solutions
$ L_{2}-$normalizable at the origin have
the following behaviour:
\be
\Psi^{(1)}(x) \mathop{\sim}_{x \rightarrow 0} \left(\begin{array}{c}
a_{1} x^{5/2} + a_{2} x^{1/2}\\ a_{1} x^{5/2} - a_{2} x^{1/2}
\end{array}\right); \quad
\Psi^{(2)}(x) \mathop{\sim}_{x \rightarrow 0}
 \left(\begin{array}{c}
b_{1} x^{1/2}\\b_{2} x^{1/2} \end{array}\right).  \label{as}
\ee
where we display explicitly the diagonalization of matrix
potential $V^{(1)}(x).$
Both potentials for
$ \delta_{3} \neq 0 $
lead to a discrete
spectrum.

The  symmetry operator $R^{(1)}$ can be calculated from
Eq.(\ref{R1}) and found
to contain also $\sigma_2$ type of terms: it
can be written as
\be
R^{(1)}=\sigma_3\partial^2 + (2i/x) \sigma_2\partial+
\{- \delta_3\cdot\sigma_1 -
(i/x^2)\cdot\sigma_2+[1/(4 x^2)-\delta_3^2 x^2/4]\cdot\sigma_3\}.
\nonumber
\ee
This operator also contains centrifugal singularities but one can
easily prove that it maps
$ L_{2}-$normalizable solutions (\ref{as}) into
themselves. Therefore it is a true symmetry operator. The partner
symmetry operator
$ R^{(2)} = \delta_{3}\sigma_{3} $ is regular.

As in Subsection 2.1. also this example allows
an interpretation in terms of the external field
as a magnetic field.
It is possible to consider a magnetic field
as a (pseudo)vector in a plane orthogonal to the one dimensional axis
 in which the particle is allowed to move $x \equiv x_2$
\footnote{This choice will allow to implement
the condition of absence of sources of magnetic field
 automatically, $\partial_i B_i = 0.$}.
Contrary to Subsection 2.1. the magnetic field in $V^{(2)}$
is homogeneous along
the $x_3$ axis, while in
$V^{(1)}$ it is {\bf not} homogeneous
(depends on $x_2$) and has non zero components
 in the $(x_1, x_3)$ plane.

To describe other examples it is useful to
introduce  $$W(x) + \widetilde W(x)
\equiv 2 F(x) = 2f_0 + 2f_i \sigma_i; \quad 2f_1 = w_1, \quad 2f_2 = w_2$$
(\ref{ladder}) becomes:
\be
2 F^{\prime} - 4 F^2 + 2\{ F , W \} = \hat \Delta . \label{delta}
\ee
The problem is now  expressed in terms of a {\it matrix non-linear
differential} equation (\ref{delta}) for $F,\, W$. Since we are unable
to present a general (analytic) discussion we provide particular
solutions of the system
(\ref{delta}) which in terms of components can be rewritten:
\be
2f_0^{\prime} - 4(f_0)^2 + 4(f_1)^2 + 4(f_2)^2 - 4(f_3)^2 + 4 f_0 w_0
+ 4 f_3 w_3 = \delta_0; \label{prime}
\ee
\be
f_1^{\prime} + 2 w_0 f_1 = 0; \label{d1}
\ee
\be
f_2^{\prime} + 2 w_0 f_2 = 0; \label{d2}
\ee
\be
2f_3^{\prime} - 8 f_0 f_3 + 4 f_0 w_3 + 4 w_0 f_3 = \delta_3. \label{d3}
\ee
The solutions presented in the following will
allow to construct explicitly the
partner potentials and the symmetry operators
by making use of the general expressions in terms only of the $f_k , \,
(k=0,1,2,3)$ and
$w_0, w_3$:
\ba
V^{(1)}(x) &=& w_0^2 + w_3^2 -w_0' +
4 (f_1^2 + f_2^2) - 4f_1'\cdot\sigma_1 -
 4f_2'\cdot\sigma_2 \nonumber\\
&+& (2w_0w_3 - w_3')\cdot\sigma_3;
\nonumber \\
V^{(2)}(x) &=& w_0^2 + w_3^2 -w_0' + 4 (f_1^2 + f_2^2 + f_0') - \delta_0
\nonumber\\
&+& \bigl[ 2w_0w_3 - w_3' + 8(2f_0f_3 - f_0w_3 -f_3w_0) + \delta_3 \bigr]
\cdot\sigma_3 .
\nonumber
\ea
It is advantageous to use not the expression (\ref{R1}) directly but to
define a symmetry operator by suitable subtraction and rescaling (see
Section 1):
\ba
\widetilde R^{(1)} &\equiv& - \frac{R^{(1)}+\delta_0h^{(1)}}{\delta_3} =
-\sigma_3\partial^2 + 4i(f_2\sigma_1-f_1\sigma_2)\partial
 \nonumber\\ &+& (2w_0w_3-
w_3') + 4(w_3f_1-iw_0f_2)\cdot\sigma_1
+ 4(w_3f_2+iw_0f_1)\cdot\sigma_2 \nonumber\\ &+&
(w_0^2+w_3^2-4f_1^2-4f_2^2-w_0')
\cdot\sigma_3 , \nonumber
\ea
and we remind that $R^{(2)} \equiv \hat\Delta .$

We now list two particular cases:
\subsection*{\normalsize Example 2. \quad $f_0
\equiv 0;\quad f_3 \not\equiv 0$}
\hspace*{1ex}
The solution is of the type
\ba
2f_1(x) &=& \gamma_1 \exp (-2 \int w_0 dx);\nonumber\\
2f_2(x) &=& \gamma_2 \exp (-2 \int w_0 dx);\nonumber\\
2f_3(x) &=& \exp (-2 \int w_0 dx) [\gamma_3 + \delta_3
\exp (+2 \int w_0 dx)] ; \nonumber\\
w_3(x) &=& (4 f_3)^{-1} \bigl[ \delta_0 - (\gamma_1^2 + \gamma_2^2)
\exp (+4 \int w_0 dx) + 4 f_3^2 \bigr]\nonumber
\ea

\subsection*{\normalsize Example 3. \quad $f_3
\equiv 0;\quad f_0 \not\equiv 0
\quad \delta_0 = 0$}
\hspace*{1ex}
The solution of the Eqs.(\ref{prime}) --- (\ref{d3})
can be written as
\ba
2f_1(x) &=& \gamma_1 \exp (-2 \int w_0 dx);\nonumber\\
2f_2(x) &=& \gamma_2 \exp (-2 \int w_0 dx);\nonumber\\
w_3(x) &=& \delta_3 / (4 f_0);\nonumber\\
f_0(x) &=& 1/2\sqrt{\gamma_1^2+\gamma_2^2}\exp (-2\int w_0dx)
\tanh \bigl(
-\sqrt{\gamma_1^2+\gamma_2^2} \int \exp (-2\int w_0dx) + C\bigr).\nonumber
\ea

\subsection*{\normalsize\bf 3.2.\quad Transparent matrix potentials}
\hspace*{1ex}
We now explore a physical case for which the Hamiltonian
$H^{(2)}$ describes free motion;
in Eqs.(\ref{eq1}), (\ref{eq2}),
(\ref{eq3}) we assume  $V^{(2)} = 0$ and,
as a consequence of SUSY,  $V^{(1)}$ becomes a so called
transparent matrix potential \cite{ais}, \cite{ci}:
\be
V^{(1)}+4F^{\prime}= 0 \label{eqn1}
\ee
\be
F^{\prime\prime}-V^{(1)}F-B^{\prime}=0 \label{eqn2}
\ee
\be
B^{\prime\prime}-V^{(1)}B=0
\label{eqn3}
\ee

These relations  allow for further simplifications by eliminating
$V^{(1)}$ and $B$ in terms of $F$ using Eq.(\ref{eqn1}) and
differentiating Eq.(\ref{eqn2}) to get $B^{\prime\prime}$ which
is then introduced in Eq.(\ref{eqn3}) leading to a linear algebraic
equation for $B$ \footnote{ Incidentally we note that
 Eq.(\ref{eqn3}) is a Schr\"odinger equation for $B$ with eigenvalue
zero. This is related to the fact that $V^{(2)}=0$
and therefore there exists a trivial zero-energy solution, namely the
constant wave function. Letting $A^+$ act on this wave function one
obtains the corresponding one for $ h^{(1)}$ with the same zero energy
as proportional to the matrix $B(x)$ acting on a constant vector.}.
Inserting back the expression for the derivative of $B$ into  Eq.(\ref{eqn2})
 we obtain
\be
F^{\prime\prime}+4F^{\prime}F+[(4 F^{\prime})^{-1}(F^{\prime\prime\prime}+
4F^{\prime 2}+4F^{\prime\prime}F)]^{\prime}=0 \label{reflectl}
\ee
or equivalently
\be
(F^{\prime\prime}+ 4F^{\prime}F)^{\prime}+4F^{\prime}\int
(F^{\prime\prime}+ 4F^{\prime}F)dx=0
\nonumber
\ee
a non-linear 4-th order equation for $F$ whose solution  may allow to
identify a class of transparent potentials in a $2\times2$
coupled channel problem.

A sufficient condition for $F$ to satisfy Eq.(\ref{reflectl}) is given
by the {\it simpler equation of second order}
\be
 (F^{\prime\prime}+ 4F^{\prime}F)=\gamma F^{\prime}\label{ioffe}
\ee
with $\gamma$ an arbitrary constant number.
It easy to verify the property that if $F(x)$ is a
 solution of the nonlinear matrix equation
(\ref{ioffe}) than $\widetilde F(x) \equiv F(x) + \alpha \cdot I$
with $\alpha = const$ is again a solution of the same equation
but for the shifted value of $\tilde \gamma = \gamma + 4\alpha .$
This peculiar property allows therefore to restrict to
the case $\gamma = 0 $ in (\ref{ioffe}).
The solution of this equation because of (\ref{eqn1})
can be searched for,  parametrizing $F(x) \equiv G(x) +
i C$ with $G(x)$ and $C$ hermitean
and $C$ a constant matrix.
We have thus to solve
\be
G^{\prime\prime}(x) + 4 G^{\prime}(x) G(x) + 4i  G^{\prime}(x) C = 0 .
\nonumber
\ee
We can expand $G$ and $C$ in Pauli matrices,
by a suitable rotation we can choose
$c_1 = c_2 = 0$ and furthermore we assume $c_0 = 0.$ Then we
arrive to a system of equations:
\ba
g_0^{\prime} + 2 g_0^2 + 2 \vec g^{\,2} &=& 2 \gamma_0;\nonumber \\
g_i^{\prime} + 4 g_0 g_i  - 4 \epsilon_{ij3} c_3 g_j &=&
2\gamma_i; \quad i,j = 1,2 ;\nonumber \\
\epsilon_{ijk}g_j^{\prime}g_k + g_0^{\prime}c_3\delta_{3i} &=& 0
; \quad i,j,k = 1,2,3 ;\nonumber \\
g_3 &=& 2\gamma_3 ,\nonumber
\ea
with constant $\gamma$'s and obvious meaning of $g$'s and $c$'s.

A solution can be found for all $\gamma_{\mu} = 0:$
\ba
g_0(x) &=& \frac{1}{4x + \beta};\nonumber \\
g_1(x) &=& g_0(x) \cos \phi(x);
\nonumber \\
 g_2(x) &=& g_0(x) \sin \phi(x);\nonumber \\
\phi (x) &\equiv& - 4 c_3 x + \zeta ,\nonumber
\ea
with $\beta$ and $\zeta$ arbitrary real constants.

By shifting
$ x \rightarrow x-\beta/4 $ and changing
$ \zeta \rightarrow \zeta+c_{3}\beta $ we obtain the transparent
hermitean matrix potential $ V^{(1)}(x) $ with a singularity
at the origin and asymptotic long range behaviour:
\ba
V^{(1)}(x)&=&\frac{1}{x^2} + \biggl[ \frac{1}{x^2} \cos \phi(x) -
\frac{4 c_3}{x} \sin \phi(x) \biggr] \cdot \sigma_1 +
 \biggl[ \frac{1}{x^2} \sin \phi(x) +
\frac{4 c_3}{x} \cos \phi(x) \biggr] \cdot \sigma_2  \nonumber \\&=&
\frac{1}{x^2}\bigl( 1 + \hat{N}(x) \bigr) -
\frac{4ic_{3}}{x}\sigma_{3}\hat{N}(x), \label{VVV}
\ea
where
$$
\hat{N}(x) \equiv \sigma_{n}\cos (4c_{3}x) +
i \sigma_{3}\sigma_{n}\sin (4c_{3}x)
$$
$$
\sigma_{n} \equiv \sigma_{1}\cos\zeta +\sigma_{2}\sin\zeta .
$$
Due to the centrifugal singularity in this potential
$$
V^{(1)}(x) \mathop{\sim}_{x \rightarrow 0}
 \frac{1}{x^{2}} \bigl( 1 + \sigma_{n} \bigr) + O(1)
$$
one should consider the scattering problem on the semiaxis
$ x \geq 0.$

To analyze the behaviour of wave functions at the origin it is
useful to choose as a basis the following matrices:
$ \sigma_{n}, \tilde\sigma_{n} \equiv ( \sigma_{2}\cos\zeta -
\sigma_{1}\sin\zeta$ ) and $ \sigma_{3} $ with non-standard
realization:
$
\sigma_{n} = \left(\begin{array}{cc}
 1 & 0 \\ 0 & -1 \end{array}\right),
\tilde\sigma_{n} = \left(\begin{array}{cc}
 0 & 1 \\ 1 & 0 \end{array}\right),
\sigma_{3} = \left(\begin{array}{cc}
 0 & -i \\ i & 0  \end{array}\right).
$
Then potential $ V^{(1)}(x) $ becomes diagonal at the origin and
its wave functions for $ x \rightarrow 0 $ are:
$$
\Psi^{(1)}(x)
\mathop{\sim}_{x \rightarrow 0}
\left(\begin{array}{c}
a \cdot x^{2}\\ b \cdot x
\end{array}\right).
$$

The dynamical symmetry operator for
$ h^{(1)} $ can be derived from
\ba
\widetilde{\widetilde{R}}^{(1)} \equiv  q^+ q^- - (h^{(1)})^2
= - 4 iC \partial^3 + \bigl( -4 G^2 -
4 C^2 - 4i [C, G] - 2 G^{\prime} \bigr)\cdot\partial^2 -
\nonumber \\
 \bigl( 8GG^{\prime} +
8iC G^{\prime} + 2G^{\prime\prime} \bigr)\partial
 - \bigl( 2G^{\prime\prime\prime} +
4G G^{\prime\prime} + 4iC G^{\prime\prime} + 16 G^{\prime 2} \bigr).
\nonumber
\ea
This operator is also singular at the origin:
\ba
&&\widetilde{\widetilde{R}}^{(1)}\mathop{\sim}_{x\rightarrow 0}
-4ic_{3}\sigma_{3}\partial^{3}+\bigl( \frac{4c_{3}}{x}\tilde\sigma_{n}
-4c_{3}^{2}+16c_{3}^{2}\sigma_{n}-32c_{3}^{3}x\tilde\sigma_{n}\bigr)
\cdot\partial^2 +
\bigl( \frac{4ic_{3}}{x^{2}}\sigma_{3} -\nonumber \\
&-&\frac{4c_{3}}{x^{2}}\tilde\sigma_{n}-32c_{3}^{3}\tilde\sigma_{n}\bigr)
\cdot\partial
+\bigl( -\frac{4ic_{3}}{x^{3}}\sigma_{3} +
\frac{4c_{3}}{x^{3}}\tilde\sigma_{n} -
\frac{12c_{3}^{2}}{x^{2}} -
\frac{12c_{3}^{2}}{x^{2}}\sigma_{n} +
\frac{32c_{3}^{3}}{x}\tilde\sigma_{n}+92c_{3}^{4}\sigma_{n} \bigr).
\nonumber
\ea
Nevertheless it maps wave functions regular at the
origin into regular ones similarly to the Example 1 (Section 3.1.).

Proceeding in the same way for $\gamma_0 > 0,
\gamma_k = 0\,\, (k=1,2,3) $ we find another solution:
\ba
g_0(x) &=& \omega \coth (4\omega x + \beta);\nonumber \\
g_1(x) &=&  \frac{\omega\cos\phi}{\sinh (4\omega x + \beta)};
\nonumber\\
 g_2(x) &=& \frac{\omega \sin \phi}{\sinh (4\omega x + \beta)};\nonumber
\ea
with
$ \phi (x) $ as before and
$ \omega \equiv \sqrt{\gamma_{0}} . $
In this case the corresponding potential
$ V^{(1)}(x) $ has the same centrifugal behaviour at the origin.
The analogous analysis of wave functions and of their transformations by
symmetry operator $ R^{(1)} $ can be performed.

One can check that $F(x)$
does not factorize the space dependence from a constant
matrix. And also we stress that
the potential $V^{(1)}(x)$ cannot be made diagonal by global
rotation and therefore
it is not to be viewed as a pair of standard
scalar reflectionless potentials: this means that there is flux from one
channel to the other.

In order to ascertain the reducible or
irreducible character of the solutions of
(\ref{ioffe})
it is important to clarify the conditions for reducibility for
the case $V^{(2)} = 0$ and $V^{(1)}(x)$ a
transparent potential.
As a consequence of (\ref{ladder}),
 (\ref{inth1}), (\ref{inth2}) these conditions read:
\be
V^{(1)} = W^2 - W^{\prime} = - 4F^{\prime};
\nonumber
\ee
\be
V^{(2)} = \widetilde W^2 + \widetilde W^{\prime} + \hat \Delta = 0;
\nonumber
\ee
\be
W^2 + W^{\prime} =  \widetilde W^2 - \widetilde W^{\prime} + \hat \Delta;
\nonumber
\ee
\be
W +  \widetilde W = 2F;
\nonumber
\ee
\be
B = W \widetilde W - \widetilde W^{\prime} = 2W F + 2F^{\prime}
\nonumber
\ee
with solution
$$ \hat \Delta = - \xi^2;\quad \widetilde W = \xi; \quad W = 2F - \xi.$$
providing as a consequence the condition $[F, F^{\prime}] = 0,$
corresponding to the factorization of $x-$dependence of the matrix
$F(x)$. It is easy to check that
our potentials $V^{(1)}(x)$
do not satisfy these conditions and therefore are not reducible.

\section*{\large\bf 4.\quad Conclusion}
\hspace*{3ex}
We have demonstrated that the SUSY approach allows to relate as SUSY
partners a
dynamical matrix system where the symmetry is manifest to another matrix
system where
this symmetry is hidden in the sense that it is not otherwise easy to
guess it (see e.g. Subsection 3.1.). Therefore one is
connecting systems with more complex dynamics
to systems which are simpler or even solvable.
For Matrix Quantum Mechanics this approach
provides examples of dynamical (matrix) systems and associated symmetry
operators: it is useful
since a straightforward general investigation of symmetries of
a dynamical systems is not an easy task. In particular the connection
between degeneracy of levels and the existence of symmetry operators
is not mandatory
and needs further clarification which presumably will
depend on detailed dynamical properties
of the system under investigation.

This line of research is not
academic but on the contrary should become a useful approach to investigate
non-trivial quantum mechanical systems.
We have restricted ourselves
to 1-dim Matrix Quantum Mechanics:
the algebraic methods we develop seem therefore
to be specifically suited for this type of dynamical systems for which spatial
symmetries (like $O(3)$) already have been used to reduce the problem to a
one dimensional (radial) problem (separation of variables).

In absence of general theorems we have studied first and second order
intertwining relations between matrix 1-dim hamiltonians
discussing reducible and irreducible transformations among them.
The examples we have discussed can be interpreted as coupled channel problems
or Pauli-type Hamiltonians and our techniques may be instrumental to their
diagonalization.
We have provided for the first time explicit examples
of irreducible transformations
in the context of transparent matrix potentials in the framework of
Matrix SUSY Quantum Mechanics.

\section*{\normalsize\bf Acknowledgements}

This work was made possible by support provided by Grant of
Russian Foundation of
Basic Researches (No. 96-01-00535) and GRACENAS Grant (No. 95-0-6.4-49).
The collaboration has been
also supported by the agreement of S.Petersburg State University, IHEP, PNPI
and INFN and by the agreement of S.Petersburg State University and
University of Bologna.

\end{document}